\documentclass[a4paper,12pt]{article}
\usepackage{verbatim} 
\usepackage{jheppub} 
                     
  \usepackage[T1]{fontenc}
\usepackage{subfig}

\usepackage[T1]{fontenc} 

\usepackage[percent]{overpic}
\usepackage{wrapfig}
\usepackage{tabu}
\usepackage{diagbox}

\usepackage{graphicx}
\usepackage{tikz}
\usepackage{import}
\usepackage{accents}
\usepackage{mathrsfs,amsmath,amssymb,slashed}
\usepackage{multirow,multicol}
\usepackage[percent]{overpic}
\usepackage{slashed}

\usepackage{wrapfig}
\usepackage{tabu}
\usepackage{diagbox}
\usepackage{mathrsfs,amsmath,amssymb,amsthm,amsfonts,tikz,graphicx,accents,hyperref,color}
\usepackage{leftidx}
\usepackage{import}
\usetikzlibrary{decorations.pathmorphing}
\DeclareFontFamily{OT1}{rsfs}{}

\DeclareFontShape{OT1}{rsfs}{m}{n}{ <-7> rsfs5 <7-10> rsfs7 <10->rsfs10}{} 

\DeclareMathAlphabet{\mycal}{OT1}{rsfs}{m}{n}

\newcommand{\nn}{\nonumber}

\newcommand{\be}{\begin{equation}}
\newcommand{\ee}{\end{equation}}
\newcommand{\pa}{\partial}

\newcommand{\bea}{\begin{eqnarray}}
\newcommand{\eea}{\end{eqnarray}}
\def\mass{\footnotesize{\textrm M}}
\def\sch{\text{\tiny Sch}}
\def\gal{\text{\tiny G} }

\DeclareMathOperator{\extdm}{d}
\newcommand{\extd}{\extdm \!}

\title{\bf {Covariant Poisson's equation in torsional Newton-Cartan gravity}}
\author[]{Mohammad Abedini$^{\, a}$,}
\author[]{Hamid R. Afshar$^{\, b}$,}
\author[]{Ahmad Ghodsi$^{\, a}$}
\affiliation[a]{Department of Physics, Faculty of Science,
Ferdowsi University of Mashhad, Mashhad, Iran}
\affiliation[b]{\it School of Physics, Institute for Research in Fundamental
Sciences (IPM),\\ P.O.Box 19395-5531, Tehran, Iran}
\emailAdd{mohammad.abedini@mail.um.ac.ir}
\emailAdd{afshar@ipm.ir}
\emailAdd{a-ghodsi@ferdowsi.um.ac.ir}

\abstract{We derive the covariant Poisson's equation of $(d+1)$-dimensional Newton-Cartan gravity with (twistless) torsion by applying the `non-relativistic conformal method' introduced in \cite{Afshar:2015aku}. We apply this method on-shell to a Schr\"odinger field theory on the  curved Newton-Hooke background. The covariance of the field equation in the presence of the non-relativistic cosmological constant, entails fixing all coefficients in the covariant Poisson's equation for (twistless) torsional Newton-Cartan gravity. We further derive Ehlers conditions and an equation associated to the torsion in this method. 
}

\begin{document}
\maketitle

\section{Introduction}
{Along with} non-AdS holography developments, non-relativistic symmetries have attracted a boost of attention recently. Although non-AdS holography is a general term which ranges from flat space to Lifshitz holography --- see \cite{Taylor:2015glc} and references therein, they all share a same feature in which the would-be holographic theory is non-Lorentzian. From this perspective it is necessary to also formulate gravity in spacetimes with non-Lorentzian structure. This formulation will be of great use for non-relativistic field theories as it provides systematic of constructing fully covariant theories by coupling non-relativistic quantum fields to background geometries. In low energy {regime} there exists many strongly coupled systems in condensed matter with Galilean symmetries such as quantum Hall states whose study is very natural from this perspective \cite{Son:2013rqa,Geracie:2016bkg}. Similar to the relativistic case where the Minkowski spacetime or other relativistic backgrounds are geometric realizations of the (Poincar\'e)-Lorentz symmetry, a geometric realization of Galilean symmetry is the so-called Newton-Cartan geometry \cite{Cartan:1923zea,Cartan:1924yea}. Apart from holographic \cite{Christensen:2013rfa,Christensen:2013lma,Hartong:2014oma} and effective field theory applications \cite{Son:2008ye,Son:2013rqa,Cho:2014vfl,Jensen:2014aia,Can:2014ota,Gromov:2014vla,Geracie:2014mta,Geracie:2016dpu} of the Newton-Cartan geometry, there have been recent connections to hydrodynamics \cite{Herzog:2008wg,Rangamani:2009xk,Brattan:2010bw,Rangamani:2008gi} and string theory \cite{Harmark:2018cdl,Harmark:2017rpg,Kluson:2018egd,Bergshoeff:2018yvt,Janiszewski:2012nf}.

Newton-Cartan gravity \cite{Cartan:1923zea,Cartan:1924yea}, a dynamical realization of this type of geometry, is a frame independent (covariant) formulation of Newtonian gravity which means it reformulates Newton's theory of gravity in a coordinate invariant way \cite{Trautman,Kuenzle:1972zw,Ehlers,Duval:1983pb,Duval:2009vt}. Newtonian gravity is quantified by the gravitational potential $\Phi$ and the mass density of the attracting massive object $\rho$ sitting together in the Poisson's equation;
\begin{align}\label{poisson}
    \triangle\Phi=4\pi G\rho\,,
\end{align}
where $\triangle=\partial_a\partial_a$ is the spatial Laplace operator and $G$ is the Newton constant. Obviously in a frame independent formulation one needs to work with the covariant version of \eqref{poisson}. Solving this equation is essential for determining the geodesic motion of a particle in the geometry. 
While a natural way of obtaining a class of non-relativistic models such as  Newton-Cartan (NC) gravity is by taking an appropriate non-relativistic limit or a fifth dimensional null reduction of the Einstein gravity, they can also be obtained in a gauging procedure from appropriate spacetime symmetry groups \cite{Frittelli:1995wr,Bergshoeff:2015uaa,VandenBleeken:2017rij,Hansen:2018ofj,Duval:1984cj,Julia:1994bs,DePietri:1994je,Andringa:2010it,Bergshoeff:2016lwr,Hartong:2015zia,Duval:1990hj,Duval:2008jg,Banerjee:2014nja,Banerjee:2015rca}.
The gauging procedure is  a powerful tool specially for constructing models of gravity which they do not seem to occupy any limiting corner of the general relativity.\footnote{A famous example is the Ho\v{r}ava-Lifshitz gravity \cite{Horava:2009uw,Horava:2008ih} which can be formulated by gauging the Schr\"odinger algebra \cite{Afshar:2015aku,Hartong:2016yrf,Devecioglu:2018apj}.} Newton-Cartan gravity has extensions to include twistless and arbitrary torsion. The presence of torsion in Newton-Cartan geometry refers to the case where a specific non-relativistic diffeomorphism is allowed along the the time coordinate implying that the time is not absolute any more. A specific type of torsion namely twistless torsion has been introduced \cite{Christensen:2013rfa,Christensen:2013lma,Geracie:2014nka} where unlike the arbitrary torsional case preserves causality \cite{Afshar:2015aku,Bergshoeff:2014uea,Bergshoeff:2017dqq,Festuccia:2016awg,Bekaert:2014bwa,banerjee:2016laq}. We will consider the effect of both twistless and arbitrary torsion to the Poisson's equation \eqref{poisson} and other Newton-Cartan equations in this work. 


In the relativistic setup the causal structure of spacetime is preserved under conformal symmetry which is a bigger symmetry group than the Poincar\'e symmetry. Conformal  symmetry is also an important feature of the AdS holography realized by the holographic boundary theory. Classical field theories invariant under global (super-)conformal group, have been exploited for constructing local (super-)Poincar\'e invariants via the so-called `conformal method' by introducing compensating multiplets and minimally coupling and then fixing them  \cite{Kaku:1978nz,Kaku:1978ea,Ferrara:1977ij,Kaku:1977pa}. The analogue of the conformal algebra in the Galilean setting is the Schr\"odinger algebra with Lifschitz scaling $z=2$ \cite{Niederer:1972zz,Jackiw:1990mb,Henkel:1993sg,Nishida:2007pj}.  In \cite{Afshar:2015aku} a similar approach denoted as the `non-relativistic conformal method' has been introduced for constructing local {Galilean} invariants in the Newton-Cartan geometry with twistless torsion by exploring and classifying classical Schr\"odinger field theories in flat background. This classification was carried out off-shell by obtaining the Ho\v rava-Lifshitz gravity  Lagrangian at $z=2$ and on-shell by obtaining the Newton-Cartan gravity  equations of motion with twistless torsion. 

In this work by focusing on the second classification, we will show that not all terms in the (twistless) torsional  Newton-Cartan field equations obtained in \cite{Afshar:2015aku} are unique. We then reapply the non-relativistic conformal method this time for a Schr\"odinger field theory on the Newton-Hooke background which is the non-relativistic version of (A)dS spacetime and re obtain the field equations and  fix the ambiguity in the Poisson's equation completely. We also manifest three maximum arbitrariness in other field equations associated to the Ehlers conditions and the torsion equation.








This paper is organized as follows. After introducing the twisted Schr\"odinger algebra which is the basis of our construction in section \ref{twistedschrodg}, we review the non-relativistic conformal construction of \cite{Afshar:2015aku} in section \ref{NCGeometry}. Then, in section \ref{nonconf} we derive the rigid twisted Schr\"odinger transformation on the Newton-Hooke background. In section \ref{TTNC} we derive field equations for twistless torsional Newton-Cartan gravity including the Ehlers conditions, Poisson's equation and the torsion equation starting from some on-shell Schr\"odinger field theories on the Newton-Hooke background. Finally in section \ref{TorNC} we relax torsion arbitrary and find the corrections to the Poisson's equation. We conclude in section \ref{Disc} and also gather related formulas for Schr\"odinger gravity in appendix \ref{App A}.

\section{Twisted Schr\"odinger algebra}\label{twistedschrodg}

The Galilean algebra is a Kinematical algebra consisting of $\{H, P_{a}, J_{ab}, G_{a} \} $ as generators of time and spatial transformations, spatial rotation and Galilean boost respectively. Accordingly they transform as scalar, vector, tensor and vector representations of $so(d)$. Apart from that the only non-trivial  commutation relation is
\begin{align}
[H,G_a]=P_{a}\,.\label{GAlg}
\end{align}
The Bargmann extension of this algebra requires to add a new generator $N$ which commutes with all generators of the algebra. The algebra then includes one more commutation relation in addition to (\ref{GAlg})
\be
[G_a,P_b]=-N \delta_{ab}\,.\label{BAlg}
\ee
The Newton-Hooke deformation of this algebra includes yet another non-trivial commutation relation,
\be\label{HPcom}
[H,P_a]=-\frac{1}{R^2} G_a\,.
\ee
As we will shortly argue, $-1/R^2$ can be denoted as a non-relativistic cosmological constant. This can be argued as follows; had we started from the (A)dS$_{d+1}$ algebra whose spacetime translation generators do not commute;
\begin{align}
    [P_a,P_b]=-\frac{1}{L^2} J_{ab}\,,\qquad\qquad[H,P_a]=-\frac{c^2}{3L^2} G_{a}\,,
\end{align}
with $-1/L^2$ being the relativistic cosmological constant, the Newton-Hooke extension of the Galilei algebra is obtained by the scaling limit \cite{Gibbons:2003rv};
\begin{align}
    c\to \infty\,,\qquad\qquad \frac{c^2}{3L^2}\to\frac{1}{R^2}\,.
\end{align}
The conformal extension of the Newton-Hooke algebra is the  Schr\"odinger algebra. 
Namely, analogues to the Bargmann algebra, the Newton-Hooke algebra can be embedded into the Schr\"odinger algebra $Sch_z$ in $d+1$ dimensions  as a subalgebra, where the dynamical exponent $z$ refers to the anisotropic scale transformation of space-time coordinates $i.e.$
\be
t\rightarrow \lambda^z t\,,\qquad x^a\rightarrow\lambda x^a\,,\qquad a=1,\cdots,d\,.
\ee
At $z=2$ the Schr\"odinger algebra includes a new generator $K$ for special conformal transformation and $N$ remains central. 
The commutation relations are given in \eqref{bbb}
where the Schr\"odinger algebra is written in a new basis which is suitable for this embedding. We may view this by the following linear change of basis (twist) in the time translation generator of the standard Schr\"odinger algebra   
\be\label{Hnew}
H\rightarrow  H-\frac{1}{R^2} K\,.
\ee
This defines an isomorphism on the algebra level which gives back \eqref{HPcom} and modifies also the following commutation relation;
\be
[D,H]=-2(H-\frac{2}{R^2} K)\,.
\ee
We sometimes refer to the Schr\"odinger algebra in this particular basis as {\it twisted Schr\"odinger algebra}.

\section{Newton-Cartan geometry}\label{NCGeometry}
The  non-relativistic structure of Newton-Cartan geometry in $d+1$ dimensions can be described as a spacetime with a non-vanishing clock one-form $\tau=\tau_\mu dx^\mu$ and a degenerate symmetric tensor $h^{\mu\nu}$  such that its temporal part is zero $\tau_\mu h^{\mu\nu}=0$. These variables define the local time direction and the inverse of a metric in  spatial directions respectively . They are related to Galilean vielbein $(e_\mu{}^0,e_\mu{}^a)$ in frame formalism  by
\begin{align}\label{tuah}
\tau_\mu \equiv e_\mu{}_0\,,\qquad\qquad h^{\mu\nu} \equiv e^\mu{}_a e^\nu{}_b\, \delta^{ab}\,,\qquad\qquad a=1, \cdots, d\,,
\end{align}
where the frame indices $a$ refer to spatial local Galilean frame while $\mu$ is the curved index. The inverse of these variables are naturally defined from the following orthonormality conditions;
\begin{alignat}{2}
\tau^\mu \tau_\mu &= 1\,, & \qquad \qquad  \tau^\mu e_\mu{}^a &= 0 \,, \nonumber \\
\tau_\mu e^\mu{}_a &= 0 \,, & \qquad \qquad e_\mu{}^a e^\mu{}_b &= \delta^a_b \,. \label{projective-inverse}
\end{alignat}
We specially find this frame formalism useful for going to the non-coordinate basis and doing calculation there. In this basis a given Newton-Cartan tensor $T_\mu$ is represented as
\begin{align}
    T_\mu=T_0\tau_\mu+T_ae_\mu{}^a\,,\qquad\quad\quad  T_0 \equiv \tau^\nu T_\nu\,,\quad \quad T_a\equiv e^\nu{}_a T_\nu\,.
\end{align} 
Another advantage of the frame formalism is its close connection with gauging viewpoint of the spacetime algebras where vielbein {behave} as gauge fields of spacetime translations. 

In these non-Lorentzian setups the generator of Galilei-boost acts non-trivially as in \eqref{GAlg}-\eqref{HPcom}. The commutation relation \eqref{GAlg} results in non-invariance of the vielbein $e_\mu{}^a$ and the inverse $\tau^\mu$ under boost;
\begin{align}
	\delta_G \tau^\mu = - \Lambda^a e^\mu{}_a\,,\hskip 2 truecm \delta_G e_\mu{}^a = \Lambda^a\tau_\mu\,,
\end{align}
where $\Lambda^a$ is the parameter of local Galilei  boost transformation. We can however remedy this by associating a vector field $M_\mu$ to the generator of central charge symmetry whose transformation under boost is dictated by the commutation relation \eqref{BAlg} as a shift in its spatial projection $\delta M_a=\Lambda_a$. This vector field is essential to define boost-invariant quantities 
\begin{align}
\hat e_\mu{}^a= e_\mu{}^a-\tau_\mu M^a\,,\qquad
\hat\tau^\mu=\tau^\mu+ e^\mu{}^a M_a\,,\qquad
\Phi=M_0+\frac{1}{2}M_a M_a\,.\label{ehat}
\end{align} 
Out of these independent variables $(\tau_\mu,e_\mu{}^a, M_\mu)$ and their inverses we can construct Bargmann connection gauge fields associated to the Galilei  rotation and boost as dependent ones whose variations under the algebra  are preserved up to some torsion terms;
 \begin{subequations}\label{depgaugefields}
 \begin{align}
  \Omega_\mu{}^a{}^b&=-2e^\nu{}^{[a} \pa_{[\mu} e_{\nu]}{}^{b]}+e_\mu{}_ce^\nu{}^ae^\rho{}^b \pa_{[\nu} e_{\rho]}{}^c-\tau_\mu e^\nu{}^ae^\rho{}^b\pa_{[\nu} M_{\rho]}\,,\label{Barg gauge fields1}\\[.2truecm]
  \Omega_\mu{}^a&=\tau^\nu \pa_{[\mu} e_{\nu]}{}^a+\tau^\nu e^\rho{}^a e_\mu{}_b \pa_{[\rho} e_{\nu]}{}^{b}+e^\nu{}^a\pa_{[\mu} M_{\nu]}+\tau_\mu \tau^\rho e^\nu{}^a\pa_{[\rho} M_{\nu]}\,.\label{Barg gauge fields2}
 \end{align}
 \end{subequations}
Curvatures associated to the boost and spatial rotation and time translation are constructed from the gauge fields in the standard way, 
\begin{subequations}\label{curvatures}
\begin{align}
\mathcal{ R}_{\mu\nu}{}{}^a(G)&=2\partial_{[\mu} \Omega_{\nu]}{}^a-2\Omega_{[\mu}{}{}^{ab}\Omega_{\nu] b}+\frac{2}{R^2}e_{[\mu}{}^a\tau_{\nu]}
\,,\\
\mathcal{R}_{\mu\nu}{}{}^{ab}(J)&=2\partial_{[\mu}\Omega_{\nu]}{}^{ab}-2\Omega_{[\mu}{}^{ca} \Omega_{\nu]}{}^{b}{}_c\,,\\[.1truecm]
\mathcal{ R}_{\mu\nu}{}{}^{}(H)&=2\partial_{[\mu} \tau_{\nu]}\,.
\end{align}
\end{subequations}
The spatial and temporal projection of the last curvature is identifying torsion in the theory.

\begin{center}
\begin{tabular}{cccccc}\label{tabletorsion}
Twistless Torsion &&&&& Arbitrary Torsion \\  
${\mathcal R}_{a0}(H)\neq0\,, {\mathcal R}_{ab}(H)=0$& &&&&
${\mathcal R}_{a0}(H)\neq0\,, {\mathcal R}_{ab}(H)\neq0 $ \\
\end{tabular}
\end{center}
The discussion on the arbitrary torsional Newton-Cartan is postponed to section \ref{TorNC}. For simplicity we use another name for ${\mathcal R}_{a0}(H)/2$ which is generically non-zero in the twistless torsional Newton-Cartan gravity;
\begin{align}
b_a\equiv\tfrac12{\mathcal R}_{a0}(H)=e_a{}^\mu\tau^\nu\partial_{[\mu}\,\tau_{\nu]}\,.
\end{align}
A natural interpretation for $b_a$ is given by the gauge field associated to dilatation in Schr\"odinger gravity --- see appendix \ref{App A}. 

Here we find it useful to give the spatial and temporal projection of gauge fields associated to rotation and boosts by substituting $M_\mu=M_0\tau_\mu+M_ae^a{}_\mu$ in \eqref{depgaugefields};
 \begin{subequations}\label{depgaugefields2}
 \begin{align}
  \Omega_0{}^a{}^b&=-2\tau^\mu e^\nu{}^{[a} \pa_{[\mu} e_{\nu]}{}^{b]}-\partial^{[a}M^{b]}+\Omega^{[ab]}{}_cM^c-\tfrac12{\mathcal R}^{ab}(H)M_0\,,\label{Barg gauge fields1 a}\\[.2truecm]
  \Omega_c{}^a{}^b&=-2e^\mu{}_ce^\nu{}^{[a} \pa_{[\mu} e_{\nu]}{}^{b]}+e^\mu{}^ae^\nu{}^b \pa_{[\mu} e_{\nu]}{}^c\,,\label{Barg gauge fields1 b}\\[.2truecm]
  \Omega_0{}_a&=2\tau^\mu e^\nu{}_a\partial_{[\mu}e_{\nu]}{}^bM_b+\partial_0M_a-\partial_aM_0-2b_aM_0\,.\label{Barg gauge fields2 a}\\[.2truecm]
  \Omega_{ab}&=e_a{}^\mu\tau^\nu \pa_{[\mu} e_{\nu]}{}_b+e^\mu{}_b\tau^\nu  \pa_{[\mu} e_{\nu]}{}_{a}+\partial_{[a}M_{b]}-\Omega_{[a b]}{}_cM^c+\tfrac12{\mathcal R}_{ab}(H)M_0\,.\label{Barg gauge fields2 b}
 \end{align}
 \end{subequations}
From equations \eqref{depgaugefields2} it is easy to prove two identities;
\begin{align}
    &{\mathcal D}_{[a}M_{b]}=\tfrac12{\mathcal R}_{ba}(H)M_0\,,\label{ident1}\\[.2truecm]
    &\mathcal{D}_0 M_a=\mathcal{D}_a\Phi + 2b_aM_0-M_b\mathcal{D}_aM_b
\,,\label{ident2}
\end{align}
where $\mathcal{D}$ is the  covariant derivative with respect to Galilean spatial rotation and boost in the absence of torsion i.e.
\begin{align}
    \mathcal{D}_\mu M^a=\partial_\mu M^a-\Omega_{\mu}{}^{ac} M_c-\Omega_{\mu}{}^{a}\,,\qquad\mathcal{D}_\mu b_a = \partial_{\mu}b_a - \Omega_{\mu a}{}^{c}b_c\,,\qquad\mathcal{D}_\mu\Phi=\partial_\mu\Phi\,.
\end{align}
Here we also define the corresponding covariant derivatives in the presence of twistless torsion;
\begin{subequations}\label{ehlerscon}
\begin{align}
K_a &= \mathcal{D}_{0} b_a + M\cdot\mathcal{D}\, b_a + b_aM \cdot b - M_a b\cdot b\,,\label{def. Galinv1}\\[.1truecm]
K_{ab} &= \mathcal{D}_a M_b +M_ab_b+M_bb_a\,,\qquad K=\delta^{ab} K_{ab}\,.\label{def. Galinv2}
\end{align}
\end{subequations}

\section{Non-relativistic conformal method}\label{nonconf}
Using the relativistic conformal method one can obtain Poincar\'e invariants such as Einstein-Hilbert term from real free conformal scalar field theories coupled to conformal gravity. The role of the scalar field is to compensate for the scale symmetry. One can explicitly break the conformal symmetry down to Poincar\'e by fixing the value of the scalar field such that its variation becomes zero. As it was described in \cite{Afshar:2015aku} and will be briefly  reviewed here, this method can be applied for obtaining Galilean invariants by starting from a  Schr\"odinger field theory and breaking the associated scaling and central charge symmetries. Since in this case there are two symmetries that should be compensated we introduce two real scalar fields $(\varphi,\chi)$;
\begin{align}\label{compensators}
    \delta\varphi=w\Lambda_D\varphi\,,\hskip 1.5truecm \delta\chi=\mass\,\sigma\,,
\end{align}
where $w$ and $\mass$ are the corresponding weights of these fields under local scaling $\Lambda_D$ and local central charge $\sigma$ symmetries. The transformation rules \eqref{compensators} automatically define the covariant derivatives associated to these fields which would be useful for coupling these fields to Schr\"odinger gravity later;
\begin{align}
    D_\mu\varphi=\left(\partial_\mu-wb_\mu\right)\varphi\,,\qquad\qquad D_\mu\chi=\partial_\mu\chi-\mass m_\mu\,.
\end{align}
For a list of higher Schr\"odinger covariant derivatives see appendix \ref{App A}.

Reminding that Schr\"odinger independent gauge fields transform under these symmetries according to \eqref{tr};
 \begin{align}\label{non-rel-trans}
  \delta (\tau_\mu)^\sch &= 2\Lambda_{\text{D}}(\tau_\mu)^{\text{\tiny Sch}}\,,\hskip 1truecm
 \delta (e_\mu{}^a)^{\text{\tiny Sch}} = \Lambda_{\text{D}}(e_\mu{}^a)^{\text{\tiny Sch}}\,, \hskip 1truecm \delta m_\mu = \partial_\mu\sigma\,,
 \end{align}
the transformation of $(\varphi,\chi)$  compensating fields in \eqref{compensators} is such that  Galilei variables $(\tau_\mu)^\gal$, $(e_\mu{}^a)^\gal$ and $M_\mu$ remain invariant under these symmetries once we relate them to the corresponding Schr\"odinger variables $(\tau_\mu)^{\sch}$, $(e_\mu{}^a)^{\sch}$ and $m_\mu$ as below;
\begin{align}\label{ddd}
	&(\tau_\mu)^\gal= \varphi ^{-\frac{2}{w}} ({\tau_\mu})^{\sch}\,,\qquad (e_\mu{}^a)^\gal = \varphi^{-\frac{1}{w}} (e_\mu{}^a)^{\sch}\,,\qquad
	M_\mu=m_\mu-\frac{1}{\mass}\partial_\mu\chi\,.
\end{align}
In principle one can start from a Galilean invariant made out of $(\tau_\mu)^\gal$, $(e_\mu{}^a)^\gal$ and $M_\mu$ variables and make it invariant under extra scaling and central charge symmetries by introducing $(\varphi,\chi)$ St\"ukelberg fields and substituting \eqref{ddd} into the Galilean invariant theory. This is the same as coupling a Schr\"odinger invariant ($\varphi,\chi$)-field theory to Schr\"odinger gravity. This implies that in principle one can also find Galilean invariants starting from a Schr\"odinger field theory of $(\varphi,\chi)$. In \cite{Afshar:2015aku} this has been done for the case of a Schr\"odinger field theory in flat background. In this work we do this on a curved (Newton-Hooke) background owning a non-relativistic cosmological constant. We think this is an important check because at the end everything are gauged and there is no more a preferred reference frame, so one can just set the cosmological constant to zero and compare the invariants. In fact as we will  see the result almost matches with \cite{Afshar:2015aku} but also contains new terms in the invariant which were missed before. 
\subsection{Rigid Schr\"odinger transformation on  Newton-Hooke background}\label{rigidtrans}
In order to write the $(\varphi,\chi)$- Schr\"odinger field theory on the Newton-Hooke spacetime,  we should first derive the corresponding rigid Schr\"odinger symmetries. 
To do this we need to fix gauge fields to their background values and require their transformations to vanish --- it is clear that all gauge fields are Schr\"odinger algebra-valued;
\begin{align}\label{aaa}
&(\tau_\mu)^\sch=\delta_{\mu 0}\,,\hskip 2. truecm (e_\mu{}^a)^\sch=\delta_\mu{}^a\,,\hskip 2. truecm m_\mu=\frac{x^2}{2R^2} \delta_{\mu 0}\,.
\end{align}
The gauge fixing conditions \eqref{aaa} correspond to the Newton-Hooke spacetime or a flat Newton-Cartan geometry with a Newton potential (see e.g. \cite{Grosvenor:2017dfs}). Using the $ U(1) $ transformation and local Galilean boosts, we can introduce new coordinates with $ m_\mu=0 $. 
Imposing the  fixing conditions \eqref{aaa} on dependent gauge fields in \eqref{dependentgaugefields}, gives
\begin{align}\label{ads}
\omega_\mu{}^a=-\frac{x^a}{R^2} \delta_{\mu 0}\,, \qquad\omega_\mu{}^{ab}=0\,,\qquad  f_\mu=0\,,\qquad b_a=0\,.
\end{align}
Requiring fixing conditions \eqref{aaa} and \eqref{ads} are preserved under the transformation rules \eqref{Atransf}-\eqref{tr},  we assume that the left hand side of \eqref{Atransf} is zero and solve for those parameters associated to global (rigid) transformations. We find,
	\begin{align}\label{par}
    	\Lambda^a&=\lambda^a \cos\frac{t}{R}+\frac{b^a}{R} \sin\frac{t}{R}\!-\!\lambda_K x^a \cos\frac{2}{R}t\!-\!\frac{2}{R}\lambda_D x^a \sin\frac{2}{R}t\,,\quad\,\,
\Lambda^{ab}=\lambda^{ab}\,,
\nn \\
	\Lambda_K&=\lambda_K \cos\frac{2}{R}t+\frac{2}{R}\lambda_D \sin\frac{2}{R}t\,,\qquad \,\,
	\Lambda_D=-\frac{R}{2}\lambda_K \sin\frac{2}{R}t+\lambda_D \cos\frac{2}{R}t\,,\nn\\
	\sigma&=\sigma_0+\frac{1}{2}\lambda_k x^2 \cos\frac{2}{R}t+\frac{1}{R} \lambda_D x^2 \sin\frac{2}{R}t-\lambda^a x^a \cos\frac{t}{R}
	-\frac{1}{R}b^a x^a \sin\frac{t}{R}\,,
\nn\\
	\xi^a&=-\lambda^{ab} \, x^b+\frac{R}{2}\lambda_K \, x^a \sin\frac{2}{R}t-\lambda_D \, x^a \cos\frac{2}{R}t-\lambda^a R \sin\frac{t}{R}
	+b^a \cos\frac{t}{R}\,,\nonumber \\
	\xi^0&=b^0+\frac{R^2}{2}\lambda_K-\frac{R^2}{2}\lambda_K \cos\frac{2}{R}t-R\lambda_D \sin\frac{2}{R}t\,,
	\end{align}
where  $ b^0 $, $ b^a $, $ \lambda^a $, $ \lambda^{ab} $, $ \lambda_K $, $ \lambda_D $ and  $ \sigma_0 $ are constant parameters. One may check that whenever $ R\rightarrow\infty $, then we can obtain the same results in \cite{Afshar:2015aku} for Bargmann algebra.
The relations between above parameters are as follows
\begin{align}\label{LLL}
-\partial_0\xi^0 &=2\Lambda_D\,,\qquad\qquad\quad -\partial_0\xi^a=-\partial_a\sigma=\Lambda^a \,, \nonumber\\
\partial_c\partial_b\xi^a &=0\,,\qquad\qquad\qquad\,\,
\partial_0\partial_b\xi^a=\delta_b{}^a \Lambda_K \,,\nonumber\\ \partial_0\partial_0\Lambda_D &=-4\Lambda_D/R^2,\quad\quad\,\, -\partial_0\Lambda_D =\Lambda_K\,.
\end{align}
We use the relations \eqref{LLL} to prove that the equations of motion are invariant under rigid or local Newton-Hooke transformations.
Finally the action of above rigid transformations on the $(\phi,\chi)$ scalar fields become,
\begin{align}\label{phitrans}
    \delta\varphi&=\left(\xi^0\partial_0+\xi^a\partial_a + w\Lambda_D\right)\varphi\nn\\[.3truecm]
    &=b^0\partial_0\varphi+T'b^a\partial_a\varphi-T\lambda^a\partial_a\varphi-\lambda^{ab}x_b\partial_a\varphi+\lambda_k\left(T^2\partial_0\varphi+TT'x^a\partial_a\varphi-wTT'\varphi\right)\nn\\[.3truecm]
    &-\lambda_D\left(2TT'\partial_0\varphi+[1-2T^2/R^2]x^a\partial_a\varphi- w[1-2T^2/R^2]\varphi\right)\,,\\[.3truecm]
  \label{chitrans}    
    \delta\chi&=\left(\xi^0\partial_0+\xi^a\partial_a\right)\chi +\mass\sigma\nn\\[.3truecm]
    &= b^0\partial_0\chi+b^a\left(T'\partial_a\chi-\mass x_aT/R^2\right)-\lambda^a\left(T\partial_a\chi+\mass x_aT'\right)
    -\lambda^{ab}x_b\partial_a\chi\nn\\[.3truecm]
    &+\lambda_k\left(T^2\partial_0\chi+TT'x^a\partial_a\chi+\mass x^ax_a[1-2T^2/R^2]/2\right)\nn\\[.3truecm]
    &-\lambda_D\left(2TT'\partial_0\chi+[1-2T^2/R^2]x^a\partial_a\chi- 2\mass x^ax_aTT'/R^2\right)\,,
\end{align}
where $T=R\sin \frac{t}{R}$ and prime is the derivative with respect to $t$. It is obvious that when $R\to\infty$ we recover $T\to t$ and  the transformation rules in \eqref{phitrans} and \eqref{chitrans} match with those given in \cite{Afshar:2015aku}  on the flat background.  

\section{Twistless torsional Newton-Cartan gravity equations}\label{TTNC}
Newton-Cartan field equations are covariantly constructed from curvature 2-forms \eqref{curvatures} and should be invariant under all possible Newton-Cartan transformations including boost symmetry. Since we have already exhausted some of the curvature 2-forms to solve dependent gauge fields \eqref{depgaugefields} in terms of independent gauge fields $(\tau_\mu, e_\mu{}^a,M_\mu)$ we are left with few other possibilities. In the torsionless case it is known that the Newton-Cartan covariant field equations of motion can be expressed in terms of the following set of curvature 2-forms in the Bargmann algebra \cite{Andringa:2010it};
\begin{subequations}\label{NCGeq}
\begin{align}\label{eee}
\mathcal{R}_{\mu\nu}(H)&=0\,,\\[.2truecm]
\tau^\mu e^\nu{}_a\mathcal{ R}_{\mu\nu}{}{}^a(G)&=0\,,\label{www}\\[.3truecm]
e^\nu{}_c\mathcal{R}_{\mu \nu}{}{}^{ca}(J)&=0\,.\label{rot}
\end{align}
\end{subequations}
In fact it can be easily shown that the same set of equations are also invariant under boost and work for the Newton-Hooke algebra --- see also \cite{ZOJER:2016aoj}. We drop the label $\gal$ from quantities above as it is understood that they are Galilean variables which should not be confused with their Schr\"odinger partners in appendix \ref{App A}. 

The equation \eqref{eee} is the torsionless condition  and it endows the Newton-Cartan geometry with a foliation with respect to the absolute time which means that the clock 1-form $\tau$ is closed $\extd\tau=0$. The second condition \eqref{www} gives the covariant version of the Poisson's equation  \cite{Andringa:2013mma} and the third equation \eqref{rot} corresponds to Ehlers conditions. These conditions are all invariant once torsion is zero and provide  $\frac{d(d+1)}{2}+d+1$ equations for the same number of unknown variables; the spatial metric $\gamma_{ij}$, the shift vector $N_i$  and  the Newton potential $\Phi$  defined in \eqref{ehat}.  Note that due to the zero torsion condition \eqref{eee}, the laps function $N$ is fixed to unity. This is no more the case in the torsional case.
\subsection{Ehlers conditions}
In order to generalize  Ehlers conditions  in \eqref{rot} to the case of twistless torsional Newton-Cartan gravity, due to the presence of torsion $b_a$, we need to add more terms  to make it covariant under all transformations.  
We can obtain alternative equation by starting from the following Schr\"odinger invariant equations and reduce them to Galilei invariants by gauge fixing \'a la \cite{Afshar:2015aku}. Note that below $R(J)$ and $D$ are curvature 2-form and covariant derivative defined in terms of Schr\"odinger gauge fields --- see appendix \ref{App A};
\begin{subequations}\label{ehlerscond1}
\begin{align}
 R_{a b}(J)-\#_1\varphi^{-1}D_aD_b\varphi -\#_2\varphi^{-2}D_a\varphi D_b\varphi&= 0\,,\label{NC2}\\[.2truecm]
 R_{0 a}(J)+\#_1\varphi^{-1}D_0D_a\varphi +\#_2\varphi^{-2}D_0\varphi D_a\varphi+\frac{\#_2}{\mass}\varphi^{-1}(D_aD_b\chi)(D_b\varphi)&= 0\,,\label{NC1}
\end{align}
\end{subequations}
where $\#_1$ and $\#_2$ are two  arbitrary 
coefficients (at least  at this stage). Note that these extra terms were chose to zero in \cite{Afshar:2015aku} but we can add them as they transform covariantly the same. Moreover we have  defined 
\begin{align}\label{defin3}
    R_{ab}(J)=e^\mu{}_ae^\nu{}_c R_{\mu\nu}{}^{c}{}_b(J)\,,\qquad\qquad R_{0a}(J)=\tau^\mu e^\nu{}_c R_{\mu\nu}{}^{c}{}_a(J)\,.
\end{align}
Now we use equations \eqref{dependentgaugefields}-\eqref{covar2} and gauge fix $\varphi=1$ and $\chi=0$
to reduce above equations in terms of Newton-Cartan curvature 2-form $\mathcal{R}_{\mu \nu}{}{}^{ab}(J)$. This renders the torsional extension of \eqref{rot} field equations;
\begin{subequations}\label{torsionalNCGRJ}
\begin{align}
\mathcal{R}_{ab}(J) + (d-2+\#_1)\mathcal{D}_ab_b +(d-2-\#_2) b_ab_b+ \delta_{ab}\left(\mathcal{D}\cdot b -(d-2+\#_1)b\cdot b\right) &= 0\,,\label{eq: NCeom Gal inv3}
\\[.3truecm]\mathcal{D}^bK_{ab} -\mathcal{D}_aK +(1+\#_2) K_{ab}b^b -b_aK+(d-1-\#_1)K_a &= 0\,و \label{eq: NCeom Gal inv2}
\end{align}
\end{subequations}
Above we used the boost invariant covariant derivatives \eqref{ehlerscon}.
We may fix  arbitrary coefficients in equations \eqref{torsionalNCGRJ} by comparing it with a similar result in \cite{VandenBleeken:2017rij}. In four dimensions ($d=3$) the similar equation of \cite{VandenBleeken:2017rij} is;
\begin{align}
\mathcal{R}_{ac}{}^{c}{}_b(J) - 2\mathcal{D}_ab_b -4 b_ab_b= 0\,.\label{dieter}
\end{align}
This suggests that we may reproduce \eqref{dieter} form \eqref{torsionalNCGRJ} by  fixing the arbitrary coefficients as $\#_1=-3$ and $\#_2=5$ in four dimensions. We will see that for these coefficients the last contribution in \eqref{eq: NCeom Gal inv3} can be made zero by the torsion equation such that both results match.

Equations \eqref{eq: NCeom Gal inv3} and \eqref{eq: NCeom Gal inv2}  are $\frac{d(d+1)}{2}+d$ equations for $\frac{(d+2)(d+1)}{2}+1$ geometric data (corresponding to the spatial metric on constant time surfaces $\gamma_{ij}$, the lapse function $N$, the shift vector $N_i$ and the Newton potential $\Phi$).
So we should find two more equations; one is the covariant Poisson's equation for $\Phi$ and one would be the torsion equation corresponding to $N$. In the following we use non-relativistic conformal method discussed in section \ref{nonconf} to obtain these equations in the twistless torsional $b_a\neq0$ case in section \ref{poissoneq} and \ref{to}. In section \ref{TorNC} we also extend the Poisson's equation for the case of arbitrary torsion.

\subsection{Poisson's equation}\label{poissoneq}
Upon substituting Galilean variables from \eqref{ddd} into \eqref{eee} and fixing the Schr\"odinger background as in \eqref{aaa}, we obtain the following Schr\"odinger invariant constraint and equation;
\begin{align}\label{rrr}
\partial_a\varphi=0\,,\qquad\quad
\varphi^3 \partial_0\partial_0\varphi+\frac{1}{R^2}\varphi^4-\frac{1}{R^2}=0\,,\qquad\quad w =1 \,.
\end{align}
Starting from this set of equations we can 
couple them to Schr\"odinger gravity by improving partial derivatives to covariant derivatives;
\begin{align}\label{sdf}
D_a\varphi=0\,,\qquad\quad\varphi^3 D_0D_0\varphi+\frac{1}{R^2}\varphi^4-\frac{1}{R^2}=0\,,\qquad\quad w =1\,.
\end{align}
The equations \eqref{sdf}  are invariant  under all local Schr\"odinger transformation in \eqref{compensators} and \eqref{tr} when $ w=1 $. 
Finally as expected by fixing the value of the scalar field to $\varphi=1$ in \eqref{sdf} we recover the torsionless Newton-Hooke equations \eqref{eee};
\begin{align}\label{torsionlesspoisson}
b_a=0\,, \hskip 2 truecm 
\hat\tau^\mu \partial_\mu K + K^{ab}K_{ab}    -\triangle \Phi+\frac{d}{R^2}=0\,,
\end{align}
where we have used the definition \eqref{def. Galinv2} to write it in terms of boost invariant variables. Equation \eqref{torsionlesspoisson} is exactly the same equation that arises in \cite{Hansen:2018ofj} for $b_a=0$.

\paragraph{Twistless torsional.} In the presence of twistless torsion i.e. $ b_a\neq0$ the equations of motion are given in \cite{Afshar:2015aku,VandenBleeken:2017rij,Hansen:2018ofj}. 
In \cite{Afshar:2015aku} these equations were obtained by exploiting the non-relativistic conformal method starting from a Schr\"odinger field theory in flat spacetime. Here  we revisit this construction by  starting from a Schr\"odinger field theory in a curved background (Newton-Hooke spacetime) and obtain new background independent contribution to equations of motion of \cite{Afshar:2015aku}. 
In this case $ b_a $ is no longer zero which at the level of the field theory it implies $\partial_a\varphi\neq0$ and so the field equation \eqref{rrr} is not invariant under the Galilean boost, dilatation and spatial translation;
\begin{align}
\delta\Big(\varphi^3\partial_0\partial_0\varphi+\frac{1}{R^2}\varphi^4-\frac{1}{R^2}\Big)=-2\varphi^3\Lambda^a(\partial_0\partial_a\varphi)-(\partial_0\Lambda^a) \varphi^3(\partial_a\varphi)\neq0\,,
\end{align}
with $\Lambda^a$ given in \eqref{par}. In the flat limit $R\to\infty$ it was shown in \cite{Afshar:2015aku} that in order to make $\partial_0^2\varphi=0$ invariant, it is enough to consider the contribution of a compensating scalar field $\chi$ which shift-transforms under local central charge Schr\"odinger symmetry \eqref{compensators} in the following equation;
\begin{align}\label{HHH0}
\partial_0\partial_0\varphi-\frac{2}{\mass}(\partial_0\partial_a\varphi)(\partial_a\chi)+\frac{1}{\mass^2}(\partial_a\partial_b\varphi)(\partial_a\chi)(\partial_b\chi)=0\,.
\end{align}
However this equation is no more Schr\"odinger invariant in our fixed curved Newton-Hooke background where the set of all possible transformations leaving this background intact (rigid transformations) are given in \eqref{phitrans} and \eqref{chitrans}.
We may especially notice that the spatial and temporal derivatives of $\chi $ shift-transform under boost as: 
\begin{align}
\delta_{G}\partial_a\chi\sim -\mass \,\lambda_a\,,\hskip 2 truecm 
&\delta_{G}\partial_0\chi\sim \mass\, T/R^2\,x^a \lambda_a\,,
\end{align}
where $T=R\sin\frac{t}{R}$. In the flat case the temporal derivative of $\chi$ does not shift-transform under boost while here it does. The reason is that, the term which is responsible for such a boost transformation involves time dependence whose contribution to this transformation would be $T''\sim T/R^2$. In the flat spacetime limit $R\to\infty$ this is zero. As a consequence of this fact, in the flat case at this order in time and spatial derivatives, in principle we had four more possible invariants that could be added to equation \eqref{HHH0} with arbitrary coefficients
\begin{subequations}\label{otherinvs}
\begin{align}
\partial_a\varphi\partial_a&\Big(\partial_0\chi-\frac{1}{2\mass}\partial_b\chi\partial_b\chi\Big)\,,\\
\varphi^{-1}\partial_a\varphi\partial_a\varphi&\Big(\partial_0\chi-\frac{1}{2\mass}\partial_b\chi\partial_b\chi\Big)\,,\\
\partial_a\partial_a\varphi&\Big(\partial_0\chi-\frac{1}{2\mass}\partial_b\chi\partial_b\chi\Big)\,,
\\
\varphi\partial_a\partial_a&\Big(\partial_0\chi-\frac{1}{2\mass}\partial_b\chi\partial_b\chi\Big)\,.
\end{align}
\end{subequations}
This obviously results in arbitrariness of coefficients of some of invariant terms such as
$\Phi b\cdot b$ and $\Phi{\mathcal D}\cdot b$ and $b\cdot\mathcal{D}\Phi$ in the final covariant Poisson's equation \eqref{torsionlesspoisson} as they could be generated independently from the above invariants \eqref{otherinvs}. This is in fact not the case in our model where we have a generalized set of rigid Schr\"odinger transformations \eqref{phitrans} and \eqref{chitrans}  which are not all preserved by \eqref{otherinvs}. In other words none of \eqref{otherinvs} and neither any linear combination of them transform covariantly under the twisted rigid Schr\"odinger transformation \eqref{phitrans} and \eqref{chitrans}. This is  good  as it leaves us with only a single invariant made out of $\partial_0^2\varphi$ and consequently a single covariant Poisson's equation in the twistless torsional Newton-Cartan gravity as will follow below. 

To this end, we can show that in order to make the field equation \eqref{rrr} invariant under twisted rigid Schr\"odinger transformation  \eqref{phitrans} and \eqref{chitrans} in the presence of torsion $\partial_a\varphi\neq0$, we must consider combinations of first and second order derivatives $ \varphi $, $ \chi $ which are added to equation \eqref{rrr}. We will have:
\begin{align}\label{HHH}
\varphi^3\Big[&\partial_0\partial_0\varphi-\frac{2}{\mass}(\partial_0\partial_a\varphi)(\partial_a\chi)+\frac{1}{\mass^2}(\partial_a\partial_b\varphi)(\partial_a\chi)(\partial_b\chi)\nn\\&-\frac{1}{\mass}(\partial_a\partial_0\chi)(\partial_a\varphi)+\frac{1}{\mass^2}(\partial_a\partial_b\chi)(\partial_b\varphi)(\partial_a\chi)\Big]+\frac{1}{R^2}\varphi^4-\frac{1}{R^2}=0\,.
\end{align}
As was mentioned the first and the second line in \eqref{HHH} are two independent invariants in flat spacetime 
while, the new feature of considering this field theory in a curved background is to fix arbitrariness in the coefficients of these two terms in a unique way independently of the cosmological constant $-\frac{1}{R^2}$ and at the same time excluding adding any other term at this order in derivatives to \eqref{HHH}.
This implies that the final gravity field equations that we are shortly going to obtain is unique. The independence of the coefficient of the second line to the non-relativistic cosmological constant in \eqref{HHH} means that in  the final Poisson's equation there exist new terms in addition to the ones obtained before in \cite{Afshar:2015aku} even if we turn off the cosmological constant.

After gauging the field equation \eqref{HHH}, we can get the extra equation of motion of Newton-Cartan gravity by imposing the fixing conditions $ (\varphi=1, \chi=0) $ which is the twistless torsional extension of the Newton-Cartan equations of motion \eqref{torsionlesspoisson} as follows:
\begin{align}\label{EOM}
&\tau^\mu\Big(\mathcal{ R}_{\mu a}{}{}^a(G)+2M^b \mathcal{R}_{\mu b}{}{}^a{}_b(J)\Big)+M^b M^c \mathcal{R}_{ba}{}{}^a{}_c(J)-2M^a K_a+\mathcal{D}_a b_a M^b M^b \nonumber\\
&+2\Omega_\mu{}^a(-\tau^\mu b_a+b_b e^\mu{}_b M^a-e^\mu{}_a b_b M^b)\underbrace{-(M^a M^a)(b.b)-b_a \mathcal{D}_0 M_a-b_a M_b \mathcal{D}_a M_b}_{-b\cdot\left({\mathcal D}+2 b\right)\Phi}=0\,,
\end{align}
where $\mathcal{D} $ is the Galilean covariant derivative. The last three terms in \eqref{EOM} are new contribution and other terms were already derived in \cite{Afshar:2015aku}. By using the  identity \eqref{ident2},
one can show that the extra three terms in \eqref{EOM} add up to $-b\cdot\left({\mathcal D}+2 b\right)\Phi$ and consequently modify the coefficients of some of the  terms in the final equation.
We may rewrite the whole equation in terms of Newton-Cartan variables as;
\begin{align}
    \hat\tau^\mu \partial_\mu K + K^{ab}K_{ab} - \triangle \Phi - 6\,\Phi\, b\cdot b - 2\,\Phi\, \mathcal{D}\cdot b - 5\,b\cdot\mathcal{D}\Phi = -\frac{d}{R^2}\,, \label{eq: NCeom Gal inv b}
\end{align}

We can go back to the torsionless equation of motion \eqref{torsionlesspoisson}, by replacing $b_a=0 $ in the equations of motion \eqref{eq: NCeom Gal inv b}. 
In this derivation we started with the Schr\"odinger theory \eqref{HHH} which has two time derivatives and other type of derivative terms are generated by requiring covariance. 
\subsection{The torsion equation}\label{to}
In the twistless torsional Newton-Cartan case we still need one extra equation for torsion. Inspired by \cite{VandenBleeken:2017rij} we propose the following Schr\"odinger invariant field equation;
\begin{align}\label{ddphi}
    \partial_a\partial_a\varphi +\#_3\varphi^{-1}\partial_a\varphi\partial_a\varphi=0\,.
\end{align}
 Note that we have required $\partial_a\varphi\neq0$ for having a non-vanishing torsion $b_a\neq0$ and $\partial_a\partial_b\varphi\neq0$ for making the field equation associated to $\partial_0^2\varphi$ in \eqref{HHH} invariant. The field equation \eqref{ddphi} is a possibility that does not affect these requirements  and is obviously invariant under all Schr\"odinger transformation \eqref{phitrans} and \eqref{chitrans} \footnote{Both terms in \eqref{phitrans} and \eqref{chitrans}  transform the same under Schr\"odinger transformations.}. After gauging this equation by coupling to the Schr\"odinger gravity and then fixing the scalar value to unity we are left with the following extra field equation for the twistless torsional Newton-Cartan gravity;
\begin{align}\label{extraeq}
    {\mathcal D}\cdot b-(d-1-\#_3)b\cdot b=0\,.
\end{align}
Essentially  the coefficient in front of $b\cdot b$ in equation \eqref{extraeq} is arbitrary and the consequence of   adding the term $\varphi^{-1}\partial_a\varphi\partial_a\varphi$ to \eqref{ddphi} with an arbitrary coefficient. In references \cite{VandenBleeken:2017rij,Hansen:2018ofj} an equation for torsion is given in four spcetime dimensions ($d=3$) as 
\begin{align}\label{extraeq2}
    {\mathcal D}\cdot b+2b\cdot b=e^{-1}\partial_\mu\left(e\,e^\mu{}_ab^a\right)=0\,,
\end{align}
that suggests to fix this arbitrariness to be $\#_3=4$ in four spacetime dimensions such that equation \eqref{extraeq2} is reproduced.

\section{Torsional Newton-Cartan gravity equations}\label{TorNC}
 So far we have only discussed twistless torsional Newton-Cartan gravity in which  the following curvature is always assumed to be zero,
 \begin{align}
   & {\mathcal R}_{ab}(H)=2e^\mu_ae^\nu_b\partial_{[\mu}\tau_{\nu]}\,.
\end{align} 
 In this section we are willing to relax this condition and obtain the covariant Poisson's equation in the presence of arbitrary torsion ${\mathcal R}_{ab}(H)\neq0$ in three dimensions ($d=2$). We again follow the non-relativistic conformal method described in previous sections but this time for arbitrary torsion \cite{Bergshoeff:2017dqq}. Before starting this procedure it is necessary to revisit the invariance of the conventional constraints of Schr\"odinger gravity for the case that we have arbitrary torsion ${\mathcal R}_{ab}(H)\neq0$ and perhaps to  modify them such that the ultimate dependent gauge fields $(\omega_\mu{}^a,\omega_\mu{}^{ab},b_a,f_\mu)$ transform covariantly under gauge transformation. This work has been carried out in \cite{Bergshoeff:2017dqq} where it is shown that in the presence of torsion only the following constraints are modified to restore boost invariance comparing to those of Schr\"odinger gravity in the absence of torsion in \eqref{cur};
\begin{subequations}\label{cur}
\begin{align}
&R_{\mu\nu}{}{}^a(P)+R^a{}_{[\mu }(H)M_{\nu]}=0\,,\\[.25truecm]
& R_{0a}{}{}^a(G)+2M^b R_{0}{}_b(J)+M^b M^c R_{b}{}_c(J) - 2M\cdot M M_c D_a R^{ac}(H)=0\,,\label{f}
\end{align}
\end{subequations}
where $D_a R^{ac}(H)$ is defined according to \eqref{covder} while  $R_{ab}(J)$ and $M_\mu=-\frac{1}{\mass}D_\mu\chi$ being the contracted curvature 2-form and the covariant vector field as defined in \eqref{defin3} and \eqref{ddd} respectively. As a consequence of solving these constraints we find that two of those dependent gauge fields in \eqref{dependentgaugefields} get modified;
\begin{subequations}\label{depgaugetor}
\begin{align}
    &\omega_\mu{}^{ab}\to\omega_\mu{}^{ab}+2M_\mu R^{ab}(H)\,,\\[.25truecm]
    &f_0\to f_0+\frac{1}{d}M\cdot M M_c D_a R_{ca}(H)\,.
\end{align}
\end{subequations}
 Again in order to obtain the covariant Poisson's equation in the presence of arbitrary torsion we may start from the Schr\"odinger field equation \eqref{HHH} in the Newton-Hooke background \eqref{aaa}-\eqref{ads}\footnote{The torsion does not contribute to the value of dependent gauge fields in \eqref{ads}.}. After coupling the equation \eqref{HHH} to Schr\"odinger gravity by replacing derivatives with covariant ones it is no longer invariant under local Schr\"odinger transformations. 
We rewrite the covariant version of equation \eqref{HHH} as follows;
\begin{align}\label{asas}
\varphi^3\Big[&D_0D_0\varphi-\frac{2}{\mass}(D_0D_a\varphi)(D_a\chi)+\frac{1}{\mass^2}(D_aD_b\varphi)(D_a\chi)(D_b\chi)-\frac{1}{\mass}(D_0 D_a\chi)(D_a\varphi)\nonumber\\
&+\frac{1}{\mass^2}(D_aD_b\chi)(D_a\chi)(D_b\varphi)\Big]+\frac{1}{R^2}\varphi^4-\frac{1}{R^2}=0\,.
\end{align}
It is important to note that due to the presence of the torsion, in this case the spatial covariant derivatives no longer commute on scalar fields;
\begin{align}
    [D_a,D_b]\varphi=-R_{ab}(H)D_0\varphi-wR_{ab}(D)\varphi\,,\qquad\qquad [D_a,D_b]\chi=-R_{ab}(H)D_0\chi\,.
\end{align}
The variation of \eqref{asas} under local Schr\"odinger boost transformation gives;
\begin{align}\label{46}
-\frac{\varphi^4}{2\mass}\Lambda^b R^{ab}(H)(D_0 D_a\chi)+\frac{\varphi^4}{2\mass^2}\Lambda^cR^{bc}(H)(D_aD_b\chi)(D_a\chi)\,.
\end{align}
Adding extra terms to \eqref{asas} can compensate for this variation so that one can make the whole equation invariant. We manage to accomplish this in three spacetime dimensions ($d=2$). Here we present the final result;
\begin{align}\label{50}
\varphi^3\Big[&D_0D_0\varphi-\frac{2}{\mass}(D_0D_a\varphi)(D_a\chi)+\frac{1}{\mass^2}(D_aD_b\varphi)(D_a\chi)(D_b\chi)-\frac{1}{\mass}(D_0D_a\chi)(D_a\varphi)\nn\\&+\frac{1}{\mass^2}(D_aD_b\chi)(D_a\chi)-\frac{\varphi}{2\mass^2}(D_0D_a\chi)(D_b\chi)R_{ab}+\frac{\varphi}{2\mass^3}(D_aD_b\chi)(D_a\chi)(D_c\chi)R_{bc}\nn\\&-\frac{3\varphi}{16\mass^4}R_{ba}R_{bc}(D_d\chi)(D_d\chi)(D_a\chi)(D_c\chi)\Big]+\frac{1}{R^2}\varphi^4-\frac{1}{R^2}=0\,.
\end{align}
A similar calculation has been done in \cite{Bergshoeff:2017dqq} for the field equation \eqref{HHH0} in three spacetime dimensions $(d=2)$. Here we did this for the improved field equation \eqref{HHH}. In fact one feature of \eqref{HHH} is that it automatically cancels the annoying  variation terms one would get in local variation of  \eqref{HHH0} but also introduces new terms in the variation as was shown in \eqref{46}. 
Now by fixing $ \varphi=1 $ and $\chi=0  $ in \eqref{50} we recover the equation \eqref{EOM} plus new terms:
\begin{align}\label{5555}
&\text{\eqref{EOM}}-M_b e_a{}^\nu(\tau^\mu+\frac{1}{2}M^d e_d{}^\mu)(2D_{[\mu}H_{\nu]}{}^{ab}-2H_{[\mu}{}^{ac}H_{\nu]c}{}{}^b)-\frac{1}{2}M_b(\mathcal{D}_0M_a)R^{ab}(H)\nonumber\\
&-\frac{1}{2}(M^a M_c)(\mathcal{D}_a M_b)R^{bc}(H)-\frac{3}{16}M_a M^a M^c M_b R_{cd}(H)R^{bd}(H)=0\,,
\end{align}
where $ H_\mu{}^{ab} $ is given by
\begin{align}
H_\mu{}^{ab}=-\tfrac{1}{2}M_\mu R^{ab}(H)-(e_\mu{}^c M^{[a}+M^c e_\mu^{[a})R_c{}^{b]}(H)-\tau_\mu M_c M^{[a}R^{b]c}(H)\,.
\end{align}

\section{Summary and discussion}\label{Disc}
In this paper we derived the twistless  and arbitrary torsional Newton-Cartan equations of motion by applying the non-relativistic conformal method of \cite{Afshar:2015aku} in the presence of the non-relativistic cosmological constant (Newton-Hooke background). The on-shell non-relativistic conformal approach to Newton-Cartan gravity in \cite{Afshar:2015aku} has been very successful for constructing Galilean invariants. Especially for deriving the covariant Poisson's equation in this approach, one starts from a second-order-in-time derivative Schr\"odinger-covariant equation consisting of two real scalar fields $(\phi,\chi)$ compensating for non-relativistic scale and central charge symmetries. In the torsion free case the corresponding field equation is simply given by
\begin{align}\label{schr1}
    \partial_0^2\varphi=0\,,\qquad\qquad\partial_a\varphi=0\,.
\end{align}

However, due to the on-shell derivation, here we showed that in the twistless torsional case, there are some arbitrariness in the final frame independent Newton-Cartan gravity equations including the Poisson's equation. This arbitrariness is due to the presence of other possible Schr\"odinger invariant equations in the flat background \eqref{otherinvs}. In this sense some part of the equations of motion obtained in \cite{Afshar:2015aku} for Newton-Cartan gravity are not unique, as we
can add more terms to them. 
We showed that one can fix this arbitrariness by considering the Schr\"odinger field theory in a curved Newton-Hooke background where the former existing invariants are no longer invariant in this background. In fact  one needs to add new terms to the equation associated to the Poisson's equation to make it invariant. 
In the twistless torsional case we showed that equation \eqref{schr1} should be replaced by \eqref{HHH} which indeed, has a same structure as equation \eqref{schr1} if we write it in a compact form as; 
\begin{align}\label{schr2}
    \hat{\partial}_0^2\varphi=0\,,\qquad\qquad\partial_a\varphi\neq0\,,
\end{align}
with $\hat{\partial}_0\varphi=\partial_0\varphi-\frac{1}{\mass}\partial_a\chi\partial_a\varphi$.\footnote{For simplicity we set the cosmological constant to zero.}

By coupling this equation to Schr\"odinger gravity and turning off the compensator fields we obtained the Poisson's covariant equation \eqref{eq: NCeom Gal inv b}. This affects the final covariant Poisson's equation by modifying the coefficients of some terms which are non-zero in the twistless torsional case. Since the final theory is obviously background independent, the result should be independent of the chosen background for the Schr\"odinger field theory. 
In the next step we relaxed the torsion completely arbitrary and could obtain the covariant Poisson's equation using the same method in $2+1$ dimensions \eqref{5555}. 

We also improved the Ehlers conditions \eqref{torsionalNCGRJ} and also presented a new equation associated to the torsion \eqref{extraeq} in the twistless torsional case. Still there are three  arbitrary free parameters in these equations that should be fixed. Eventually our aim is to promote this construction to an off-shell level and hopefully remove such ambiguities \cite{work-in-progress}.

It would be also interesting to obtain Newton-Cartan supergravity equations by applying the non-relativistic superconformal method applied to the supersymmetric version of these Schr\"odinger equations here. Especially for the supersymmetric Poisson's equation one should first supersymmetrize \eqref{HHH}. 

\acknowledgments
HA thanks Jelle Hartong and Dieter Van den Bleeken for useful discussion in comparing our result with the ones in references \cite{VandenBleeken:2017rij,Hansen:2018ofj}. He especially thanks Dieter Van den Bleeken for  drawing his attention to the torsion equation \eqref{extraeq2} for the first time and 
the possibility of adding extra terms in \eqref{ehlerscond1} and \eqref{ddphi}. 
HA was partially supported by the Scientific and Technological Research Council of Turkey (T\"UB\.ITAK) during the last stage of this work and he thanks Bo\u gazi\c ci University for hospitality. He would like to thank the organizers of the 2018 "Applied Newton-Cartan Geometry" workshop at  Mainz Institute for Theoretical Physics for the fruitful atmosphere and providing support during his stay when this work was in progress. 
AGh's work is supported by Ferdowsi University of Mashhad under the grant 3/46799 (1397/03/08).

\appendix
\section{From Schr\"odinger to Newton-Hooke}\label{App A}
The Newton-Hooke algebra with the non-trivial commutation relations \eqref{GAlg}-\eqref{HPcom} is a subalgebra of  the Schr\"odinger algebra whose commutation relations in an appropriate basis 
are;
\begin{align}\label{bbb}
[D, H]&=-2( H -\frac{2}{R^2}K )\,,\hskip 1. truecm[H,K]=D\,,\hskip 1.9 truecm [K,P_a]=-G_a,\nonumber\\
[D,K]&=2K\,,\hskip 3. truecm [H,G_a]=P_a\,,\hskip 1.8 truecm [D,P_a]=-P_a,\nonumber\\
[D,G_a]&=G_a\,,\hskip 3 truecm[P_a,G_b]=\delta_{ab} N\,,\hskip 1.4 truecm [H,P_a]=-\frac{1}{R^2} G_a,\nonumber\\
[J_{ab}, P_c]&= 2\delta_{c[a}P_{b]}\,, \hskip 2.2 truecm  [J_{ab}, G_c] = 2\delta_{c[a}G_{b]}\,, \hskip 0.6 truecm  [J_{ab}, J_{cd}] = 4\delta_{[a[c}\,J_{b]d]}\,.
\end{align}
If we truncate $D$ and $K$ generators we end up with the Newton-Hooke algebra with central extension. While the cosmological constant in the Newton-Hooke algebra plays the role of a deformation parameter from Bargmann algebra, at their conformal extension level it is only used to change the basis linearly by shifting the time translation generator as in \eqref{Hnew} which defines an isomorphism.

In this appendix we summarize the gauging procedure and other requisite formulas of the Schr\"odinger algebra in this very basis
which is appropriate to perform the non-relativistic conformal method discussed in the bulk of the paper.
The corresponding gauge fields and gauge parameters are given in the table \ref{table5}. 
\begin{table}
\begin{center}
      \caption{Schr\"odinger  gauge fields and parameters}
  \begin{tabular}{ l l l  l l  l l  }
    ${ H}$ & $P_a$ & $G_a$ & $J_{ab}$ & $D$ & $K$ & $N$\\     \hline\hline
    $\tau_\mu$ & $e_\mu{}^a$ & $\omega_\mu{}^a$ & $\omega_\mu{}^{ab}$ & $b_\mu$ & $f_\mu$ & $m_\mu$\\
    \hline
        $\xi^0$ & $\xi^a$ & $\Lambda^a$ & $\Lambda^{ab}$ & $\Lambda_D$ & $\Lambda_K$ & $\sigma$\\
  \end{tabular}
   \label{table5}
   \end{center}
\end{table}
Transformation rules of these gauge fields follow from the algebra according to,
\be\label{Atransf}
\delta A={\mathcal L}_\xi A +\extd\epsilon +[A,\epsilon]\,,
\ee
with $A$ and $\epsilon$ being a generic gauge field 1-form and a gauge transformation parameter while ${\mathcal L}_\xi A=\left(\xi^\nu\partial_\nu A+ A_\nu \partial_\mu \xi^\nu\right)\extd x^\mu$ is the Lie derivative 1-form. It is understood that all gauge fields transform as covariant vectors under general coordinate transformations accounted for by the Lie derivative in \eqref{Atransf} along $\xi^\mu=\xi_0\tau^\mu+\xi_a e_\mu{}^a$, so we only write transformation under the rest of generators;
 \begin{subequations}\label{tr}
\begin{align} 
\delta\tau_\mu &=2 \Lambda_D \tau_\mu\,,\\[.15truecm]
\delta e_\mu{}^a &=\Lambda^a{}_{b} e_\mu{}^b+\Lambda^a\tau_\mu+\Lambda_D e_\mu{}^a\,,\\[.15truecm]
\delta\omega_\mu{}^a &=D_\mu\Lambda^a +\Lambda^a{}_b \omega_\mu{}^b+\Lambda^a b_\mu-\Lambda_D \omega_\mu{}^a+\Lambda_K e_\mu{}^a\,,\\[.15truecm]
\delta\omega_\mu{}^{ab} &=D_\mu \Lambda^{ab},\\[.15truecm]
\delta b_{\mu} &=\partial_{\mu}\Lambda_D+\Lambda_K\tau_\mu\,,\\[.0truecm]
\delta f_\mu &=\partial_\mu \Lambda_K+2 \Lambda_K b_\mu-2\Lambda_D f_\mu-\frac{4}{R^2}\tau_\mu\Lambda_D\,,\\[.0truecm]
\delta m_\mu &=\partial_\mu\sigma+\Lambda^a e_{\mu a}\,.
\end{align}
\end{subequations}
The covariant derivative $D_\mu$ is with respect to the Schr\"odinger spatial rotation gauge field $\omega_\mu{}^{ab}$;
\begin{align}\label{covder}
    D_\mu X^{ab\cdots}=\partial_\mu X^{ab\cdots} -\omega_\mu{}^{ac}X_c^{b\cdots}-\omega_\mu{}^{bc}X^a{}_c{}^{\cdots}+\cdots\,.
\end{align}
Curvatures associated to these gauge fields are
\begin{subequations}
\begin{align}
R_{\mu\nu}(H) &=2\partial_{[\mu}\tau_{\nu]}-2b_{[\mu}\tau_{\nu]}\,,\\[.1truecm]
R_{\mu\nu}{}{}^a(P) &=2D_{[\mu}e_{\nu]}{}^a-\omega_{[\mu}{}^a\tau_{\nu]}-2b_{[\mu}e_{\nu]}{}^a\,,\\[.0truecm]
R_{\mu\nu}{}{}^a(G) &=2D_{[\mu}\omega_{\nu]}{}^a-2\omega_{[\mu}{}^a b_{\nu]}-2 f_{[\mu}e_{\nu]}{}^a+\frac{2}{R^2} e_{[\mu}{}^a \tau_{\nu]}\,,\\[.0truecm]
R_{\mu\nu}{}{}^{ab}(J) &=2D_{[\mu}\omega_{\nu]}{}^{ab}\,,\\[.1truecm]
R_{\mu\nu}(D) &=2\partial_{[\mu}b_{\nu]}-2f_{[\mu}\tau_{\nu]}\,,\\[.1truecm]
R_{\mu\nu}(K) &=2\partial_{[\mu}f_{\nu]}+4b_{[\mu}f_{\nu]}-2\Lambda b_{[\mu}\tau_{\nu]}\,,\\[.1truecm]
R_{\mu\nu}(N) &=2\partial_{[\mu}m_{\nu]}-2\omega_{[\mu}{}^a e_{\nu ]a}\,.
\end{align}
\end{subequations}
These curvatures obviously obey Bianchi identities $
\partial_{[\mu} R_{\nu\rho]}+[A_{[\mu},R_{\nu\rho]}]=0$.

Solving the following conventional curvature constraints \cite{Bergshoeff:2014uea},
\begin{subequations}\label{cur2}
\begin{align}
&R_{0a}(H)=0\,,\qquad R_{\mu\nu}{}{}^a(P)=0\,,\qquad R_{\mu\nu}(N)=0\,,\qquad
R_{0a}(D)=0\,,\\[.25truecm]
& R_{0a}{}{}^a(G)+2m^b R_{0a}{}{}^a{}{}{}_b(J)+m^b m^c R_{ba}{}{}^a{}{}{}_c(J)=0\,,\label{f2}
\end{align}
\end{subequations}
would return the dependent gauge fields $\omega_\mu{}^a{}^b$, $\omega_\mu{}^a$, $b_a$ and $f_\mu$ in terms of independent fields $\tau_\mu$, $e_\mu{}^a$, $m_\mu$ and $b_0$ such that the transformations in \eqref{tr} are preserved;
\begin{subequations}\label{dependentgaugefields}
\begin{align}
  \omega_\mu{}^a&=\tau^\nu \pa_{[\mu} e_{\nu]}{}^a+\tau^\nu e^\rho{}^a e_\mu{}_b \pa_{[\rho} e_{\nu]}{}^{b}+e^\nu{}^a\pa_{[\mu} m_{\nu]}+\tau_\mu \tau^\rho e^\nu{}^a\pa_{[\rho} m_{\nu]}+e_\mu{}^a b_0\,,\label{Omegas2}\\[.2truecm]
 \omega_\mu{}^a{}^b&=-2e^\nu{}^{[a} \pa_{[\mu} e_{\nu]}{}^{b]}+e_\mu{}_ce^\nu{}^ae^\rho{}^b \pa_{[\nu} e_{\rho]}{}^c-\tau_\mu e^\nu{}^ae^\rho{}^b\pa_{[\nu} m_{\rho]}+2e_\mu{}^{[a} b^{ b]}\,,\label{Omegas1}\\[.2truecm]
 b_a&=e_a{}^\mu\tau^\nu\pa_{[\mu}\tau_{\nu]}\label{spatialb}\,,\\[.2truecm]
 f_a &=2e_a{}^\mu\tau^\nu\partial_{[\mu} b_{\nu]}\,,\label{spatialf}\\[.05truecm]
 f_0  &=\frac{2}{d}\tau^\mu e^\nu{}_a\left(D_{[\mu}\omega_{\nu]}{}^a+b_{[\mu}\omega_{\nu]}{}^a\right)+\frac{2}{d}\left(R_{0 a}{}^{ac}(J)+\tfrac{1}{2}m^aR_{ab}{}^{bc}(J)\right)m_c-\frac{1}{R^2}\,,\label{eq:vfz=2}
\end{align}
\end{subequations}
where $d$ is the space dimensions and we used the decomposition $f_\mu =\tau_\mu f_0+e_\mu{}^af_a$.
Note that unlike in \eqref{tr}, above $f_0$ is obviously not invariant under central charge symmetry $N$. We can however promote this symmetry to a St\"ukelberg symmetry by  introducing a St\"ukelberg field $\chi$ which shift-transform under $N$ \cite{Bergshoeff:2014uea}. We then break this gauge symmetry by choosing the gauge $\chi=0$ as properly used in the  non-relativistic conformal construction of the Newton-Cartan gravity theory \cite{Afshar:2015aku} which was reviewed in section \ref{nonconf}. 
\subsection{Covariant derivatives}
In order to couple the scalar $(\varphi,\chi)$-Schr\"odinger field theories \eqref{rrr} and \eqref{HHH} to Schr\"odinger gravity  we should replace all partial derivatives with Schr\"odinger-covariant derivatives
which according to the transformation rules \eqref{tr} and \eqref{compensators} for the compensating scalar fields ($ \varphi $,$ \chi $) are defined as below;
\begin{subequations}\label{covar2}
\begin{align}\label{covar1}
D_0\varphi &=\tau^\mu(\partial_\mu-w b_\mu)\varphi\,,\\
D_a \varphi &=e^\mu{}_a(\partial_\mu-w b_\mu)\varphi\,,\\
D_0D_0\varphi &=\tau^\mu(\partial_\mu D_0 \varphi-(w-2)b_\mu D_0\varphi+\omega_\mu{}^a D_a\varphi+w f_\mu\varphi)\,,\\
D_0 D_a\varphi &=\tau^\mu(\partial_\mu D_a\varphi-(w-1)b_\mu D_a\varphi-\omega_{\mu a}{}{}^b D_b\varphi)\,,\\
D_a D_b\varphi &=e^\mu{}_a(\partial_\mu D_b\varphi-(w-1)b_\mu D_b\varphi-\omega_{\mu b}{}{}^c D_c\varphi)\,,\\
D_a\chi &=e^\mu{}_a(\partial_\mu \chi-\mass \,m_\mu)\,,\\
D_a D_0\chi &=e^\mu_a(\partial_\mu D_0\chi+\omega_{\mu }{}^b D_b\chi+2b_\mu D_0\chi)\,,\\
D_0D_a\chi &=\tau^\mu(\partial_\mu D_a\chi-\omega_{\mu a}{}{}^b D_b\chi+b_\mu D_a\chi+\mass\,\omega_\mu{}^a)\,,\\
D_aD_b\chi &=e^\mu{}_a(\partial_\mu D_b\chi-\omega_{\mu b}{}{}^c D_c\chi+b_\mu D_b\chi+\mass\, \omega_\mu{}^b)\,,\\
D_aD_bD_0\chi &=e^\mu_a(\partial_\mu D_bD_0\chi+\omega_{\mu }{}^cD_b D_c\chi+3b_\mu D_b D_0\chi-f_\mu D_b\chi)\,.
\end{align}
\end{subequations}

\bibliographystyle{fullsort.bst}
 
\bibliography{references} 

\providecommand{\href}[2]{#2}\begingroup\raggedright\begin{thebibliography}{10}

\bibitem{Afshar:2015aku}
H.~R. Afshar, E.~A. Bergshoeff, A.~Mehra, P.~Parekh, and B.~Rollier, ``{A
  Schr\"odinger approach to Newton-Cartan and Ho\v rava-Lifshitz gravities},''
  {\em JHEP} {\bf 04} (2016) 145,
\href{http://www.arXiv.org/abs/1512.06277}{{\tt 1512.06277}}.

\bibitem{Taylor:2015glc}
M.~Taylor, ``{Lifshitz holography},'' {\em Class. Quant. Grav.} {\bf 33}
  (2016), no.~3, 033001,
\href{http://www.arXiv.org/abs/1512.03554}{{\tt 1512.03554}}.

\bibitem{Son:2013rqa}
D.~T. Son, ``{Newton-Cartan Geometry and the Quantum Hall Effect},''
\href{http://www.arXiv.org/abs/1306.0638}{{\tt 1306.0638}}.

\bibitem{Geracie:2016bkg}
M.~Geracie, {\em {Galilean Geometry in Condensed Matter Systems}}.
\newblock PhD thesis, Chicago U., 2016.
\newblock
\href{http://www.arXiv.org/abs/1611.01198}{{\tt 1611.01198}}.
\newblock

\bibitem{Cartan:1923zea}
E.~Cartan, ``{Sur les vari\'et\'es \`a connexion affine et la th\'eorie de la
  relativit\'e g\'en\'eralis\'ee. (premi\`ere partie)},'' {\em Annales Sci.
  Ecole Norm. Sup.} {\bf 40} (1923)
325--412.

\bibitem{Cartan:1924yea}
E.~Cartan, ``{Sur les vari\'et\'es \`a connexion affine et la th\'eorie de la
  relativit\'e g\'en\'eralis\'ee. (premi\`ere partie) (Suite).},'' {\em Annales
  Sci. Ecole Norm. Sup.} {\bf 41} (1924)
1--25.

\bibitem{Christensen:2013rfa}
M.~H. Christensen, J.~Hartong, N.~A. Obers, and B.~Rollier, ``{Boundary
  Stress-Energy Tensor and Newton-Cartan Geometry in Lifshitz Holography},''
  {\em JHEP} {\bf 01} (2014) 057,
\href{http://www.arXiv.org/abs/1311.6471}{{\tt 1311.6471}}.

\bibitem{Christensen:2013lma}
M.~H. Christensen, J.~Hartong, N.~A. Obers, and B.~Rollier, ``{Torsional
  Newton-Cartan Geometry and Lifshitz Holography},'' {\em Phys. Rev.} {\bf D89}
  (2014) 061901,
\href{http://www.arXiv.org/abs/1311.4794}{{\tt 1311.4794}}.

\bibitem{Hartong:2014oma}
J.~Hartong, E.~Kiritsis, and N.~A. Obers, ``{Lifshitz space-times for
  Schr\"odinger holography},'' {\em Phys. Lett.} {\bf B746} (2015) 318--324,
\href{http://www.arXiv.org/abs/1409.1519}{{\tt 1409.1519}}.

\bibitem{Son:2008ye}
D.~T. Son, ``{Toward an AdS/cold atoms correspondence: A Geometric realization
  of the Schrodinger symmetry},'' {\em Phys. Rev.} {\bf D78} (2008) 046003,
\href{http://www.arXiv.org/abs/0804.3972}{{\tt 0804.3972}}.

\bibitem{Cho:2014vfl}
G.~Y. Cho, Y.~You, and E.~Fradkin, ``{Geometry of Fractional Quantum Hall
  Fluids},'' {\em Phys. Rev.} {\bf B90} (2014) 115139,
\href{http://www.arXiv.org/abs/1406.2700}{{\tt 1406.2700}}.

\bibitem{Jensen:2014aia}
K.~Jensen, ``{On the coupling of Galilean-invariant field theories to curved
  spacetime},''
\href{http://www.arXiv.org/abs/1408.6855}{{\tt 1408.6855}}.

\bibitem{Can:2014ota}
T.~Can, M.~Laskin, and P.~Wiegmann, ``{Fractional Quantum Hall Effect in a
  Curved Space: Gravitational Anomaly and Electromagnetic Response},'' {\em
  Phys. Rev. Lett.} {\bf 113} (2014) 046803,
\href{http://www.arXiv.org/abs/1402.1531}{{\tt 1402.1531}}.

\bibitem{Gromov:2014vla}
A.~Gromov and A.~G. Abanov, ``{Thermal Hall Effect and Geometry with
  Torsion},'' {\em Phys. Rev. Lett.} {\bf 114} (2015) 016802,
\href{http://www.arXiv.org/abs/1407.2908}{{\tt 1407.2908}}.

\bibitem{Geracie:2014mta}
M.~Geracie, S.~Golkar, and M.~M. Roberts, ``{Hall viscosity, spin density, and
  torsion},''
\href{http://www.arXiv.org/abs/1410.2574}{{\tt 1410.2574}}.

\bibitem{Geracie:2016dpu}
M.~Geracie, K.~Prabhu, and M.~M. Roberts, ``{Physical stress, mass, and energy
  for non-relativistic matter},'' {\em JHEP} {\bf 06} (2017) 089,
\href{http://www.arXiv.org/abs/1609.06729}{{\tt 1609.06729}}.

\bibitem{Herzog:2008wg}
C.~P. Herzog, M.~Rangamani, and S.~F. Ross, ``{Heating up Galilean
  holography},'' {\em JHEP} {\bf 11} (2008) 080,
\href{http://www.arXiv.org/abs/0807.1099}{{\tt 0807.1099}}.

\bibitem{Rangamani:2009xk}
M.~Rangamani, ``{Gravity and Hydrodynamics: Lectures on the fluid-gravity
  correspondence},'' {\em Class. Quant. Grav.} {\bf 26} (2009) 224003,
\href{http://www.arXiv.org/abs/0905.4352}{{\tt 0905.4352}}.

\bibitem{Brattan:2010bw}
D.~K. Brattan, ``{Charged, conformal non-relativistic hydrodynamics},'' {\em
  JHEP} {\bf 10} (2010) 015,
\href{http://www.arXiv.org/abs/1003.0797}{{\tt 1003.0797}}.

\bibitem{Rangamani:2008gi}
M.~Rangamani, S.~F. Ross, D.~T. Son, and E.~G. Thompson, ``{Conformal
  non-relativistic hydrodynamics from gravity},'' {\em JHEP} {\bf 01} (2009)
  075,
\href{http://www.arXiv.org/abs/0811.2049}{{\tt 0811.2049}}.

\bibitem{Harmark:2018cdl}
T.~Harmark, J.~Hartong, L.~Menculini, N.~A. Obers, and Z.~Yan, ``{Strings with
  Non-Relativistic Conformal Symmetry and Limits of the AdS/CFT
  Correspondence},'' {\em JHEP} {\bf 11} (2018) 190,
\href{http://www.arXiv.org/abs/1810.05560}{{\tt 1810.05560}}.

\bibitem{Harmark:2017rpg}
T.~Harmark, J.~Hartong, and N.~A. Obers, ``{Nonrelativistic strings and limits
  of the AdS/CFT correspondence},'' {\em Phys. Rev.} {\bf D96} (2017), no.~8,
  086019,
\href{http://www.arXiv.org/abs/1705.03535}{{\tt 1705.03535}}.

\bibitem{Kluson:2018egd}
J.~Kluso\v{n}, ``{Remark About Non-Relativistic String in Newton-Cartan
  Background and Null Reduction},'' {\em JHEP} {\bf 05} (2018) 041,
\href{http://www.arXiv.org/abs/1803.07336}{{\tt 1803.07336}}.

\bibitem{Bergshoeff:2018yvt}
E.~Bergshoeff, J.~Gomis, and Z.~Yan, ``{Nonrelativistic String Theory and
  T-Duality},'' {\em JHEP} {\bf 11} (2018) 133,
\href{http://www.arXiv.org/abs/1806.06071}{{\tt 1806.06071}}.

\bibitem{Janiszewski:2012nf}
S.~Janiszewski and A.~Karch, ``{String Theory Embeddings of Nonrelativistic
  Field Theories and Their Holographic Ho\v rava Gravity Duals},'' {\em Phys.
  Rev. Lett.} {\bf 110} (2013), no.~8, 081601,
\href{http://www.arXiv.org/abs/1211.0010}{{\tt 1211.0010}}.

\bibitem{Trautman}
A.~Trautman, ``Sur la theorie newtonienne de la gravitation,'' {\em
  Compt.~Rend.~Acad.~Sci.~Paris} {\bf 247} (1963) 617.

\bibitem{Kuenzle:1972zw}
H.~P. Kuenzle, ``{Galilei and lorentz structures on space-time - comparison of
  the corresponding geometry and physics},'' {\em Annales Poincare Phys.
  Theor.} {\bf 17} (1972)
337--362.

\bibitem{Ehlers}
J.~Ehlers, ``{\"{U}ber den Newtonschen Grenzwert},''.

\bibitem{Duval:1983pb}
C.~Duval and H.~P. Kunzle, ``{Minimal Gravitational Coupling in the Newtonian
  Theory and the Covariant Schrodinger Equation},'' {\em Gen. Rel. Grav.} {\bf
  16} (1984)
333.

\bibitem{Duval:2009vt}
C.~Duval and P.~A. Horvathy, ``{Non-relativistic conformal symmetries and
  Newton-Cartan structures},'' {\em J. Phys.} {\bf A42} (2009) 465206,
\href{http://www.arXiv.org/abs/0904.0531}{{\tt 0904.0531}}.

\bibitem{Frittelli:1995wr}
S.~Frittelli and O.~Reula, ``{On the Newtonian limit of general relativity},''
  {\em Commun. Math. Phys.} {\bf 166} (1994) 221,
\href{http://www.arXiv.org/abs/gr-qc/9506077}{{\tt gr-qc/9506077}}.

\bibitem{Bergshoeff:2015uaa}
E.~Bergshoeff, J.~Rosseel, and T.~Zojer, ``{Newton-Cartan (super)gravity as a
  non-relativistic limit},'' {\em Class. Quant. Grav.} {\bf 32} (2015), no.~20,
  205003,
\href{http://www.arXiv.org/abs/1505.02095}{{\tt 1505.02095}}.

\bibitem{VandenBleeken:2017rij}
D.~Van~den Bleeken, ``{Torsional Newton-Cartan gravity from the large c
  expansion of general relativity},'' {\em Class. Quant. Grav.} {\bf 34}
  (2017), no.~18, 185004,
\href{http://www.arXiv.org/abs/1703.03459}{{\tt 1703.03459}}.

\bibitem{Hansen:2018ofj}
D.~Hansen, J.~Hartong, and N.~A. Obers, ``{An Action Principle for Newtonian
  Gravity},'' {\em Phys. Rev. Lett.} {\bf 122} (2019), no.~6, 061106,
\href{http://www.arXiv.org/abs/1807.04765}{{\tt 1807.04765}}.

\bibitem{Duval:1984cj}
C.~Duval, G.~Burdet, H.~P. Kunzle, and M.~Perrin, ``{Bargmann Structures and
  Newton-cartan Theory},'' {\em Phys. Rev.} {\bf D31} (1985)
1841--1853.

\bibitem{Julia:1994bs}
B.~Julia and H.~Nicolai, ``{Null Killing vector dimensional reduction and
  Galilean geometrodynamics},'' {\em Nucl. Phys.} {\bf B439} (1995) 291--326,
\href{http://www.arXiv.org/abs/hep-th/9412002}{{\tt hep-th/9412002}}.

\bibitem{DePietri:1994je}
R.~De~Pietri, L.~Lusanna, and M.~Pauri, ``{Standard and generalized Newtonian
  gravities as 'gauge' theories of the extended Galilei group. I. The standard
  theory},'' {\em Class. Quant. Grav.} {\bf 12} (1995) 219--254,
\href{http://www.arXiv.org/abs/gr-qc/9405046}{{\tt gr-qc/9405046}}.

\bibitem{Andringa:2010it}
R.~Andringa, E.~Bergshoeff, S.~Panda, and M.~de~Roo, ``{Newtonian Gravity and
  the Bargmann Algebra},'' {\em Class. Quant. Grav.} {\bf 28} (2011) 105011,
\href{http://www.arXiv.org/abs/1011.1145}{{\tt 1011.1145}}.

\bibitem{Bergshoeff:2016lwr}
E.~A. Bergshoeff and J.~Rosseel, ``{Three-Dimensional Extended Bargmann
  Supergravity},'' {\em Phys. Rev. Lett.} {\bf 116} (2016), no.~25, 251601,
\href{http://www.arXiv.org/abs/1604.08042}{{\tt 1604.08042}}.

\bibitem{Hartong:2015zia}
J.~Hartong and N.~A. Obers, ``{Ho\v{r}ava-Lifshitz gravity from dynamical
  Newton-Cartan geometry},'' {\em JHEP} {\bf 07} (2015) 155,
\href{http://www.arXiv.org/abs/1504.07461}{{\tt 1504.07461}}.

\bibitem{Duval:1990hj}
C.~Duval, G.~W. Gibbons, and P.~Horvathy, ``{Celestial mechanics, conformal
  structures and gravitational waves},'' {\em Phys. Rev.} {\bf D43} (1991)
  3907--3922,
\href{http://www.arXiv.org/abs/hep-th/0512188}{{\tt hep-th/0512188}}.

\bibitem{Duval:2008jg}
C.~Duval, M.~Hassaine, and P.~A. Horvathy, ``{The Geometry of Schrodinger
  symmetry in gravity background/non-relativistic CFT},'' {\em Annals Phys.}
  {\bf 324} (2009) 1158--1167,
\href{http://www.arXiv.org/abs/0809.3128}{{\tt 0809.3128}}.

\bibitem{Banerjee:2014nja}
R.~Banerjee, A.~Mitra, and P.~Mukherjee, ``{Localization of the Galilean
  symmetry and dynamical realization of Newton-Cartan geometry},'' {\em Class.
  Quant. Grav.} {\bf 32} (2015), no.~4, 045010,
\href{http://www.arXiv.org/abs/1407.3617}{{\tt 1407.3617}}.

\bibitem{Banerjee:2015rca}
R.~Banerjee and P.~Mukherjee, ``{New approach to nonrelativistic diffeomorphism
  invariance and its applications},''
\href{http://www.arXiv.org/abs/1509.05622}{{\tt 1509.05622}}.

\bibitem{Horava:2009uw}
P.~Horava, ``{Quantum Gravity at a Lifshitz Point},'' {\em Phys. Rev.} {\bf
  D79} (2009) 084008,
\href{http://www.arXiv.org/abs/0901.3775}{{\tt 0901.3775}}.

\bibitem{Horava:2008ih}
P.~Horava, ``{Membranes at Quantum Criticality},'' {\em JHEP} {\bf 03} (2009)
  020,
\href{http://www.arXiv.org/abs/0812.4287}{{\tt 0812.4287}}.

\bibitem{Hartong:2016yrf}
J.~Hartong, Y.~Lei, and N.~A. Obers, ``{Nonrelativistic Chern-Simons theories
  and three-dimensional Ho\v rava-Lifshitz gravity},'' {\em Phys. Rev.} {\bf
  D94} (2016), no.~6, 065027,
\href{http://www.arXiv.org/abs/1604.08054}{{\tt 1604.08054}}.

\bibitem{Devecioglu:2018apj}
D.~O. Devecioglu, N.~Ozdemir, M.~Ozkan, and U.~Zorba, ``{Scale invariance in
  Newton-Cartan and Ho\v rava-Lifshitz gravity},'' {\em Class. Quant. Grav.}
  {\bf 35} (2018), no.~11, 115016,
\href{http://www.arXiv.org/abs/1801.08726}{{\tt 1801.08726}}.

\bibitem{Geracie:2014nka}
M.~Geracie, D.~T. Son, C.~Wu, and S.-F. Wu, ``{Spacetime Symmetries of the
  Quantum Hall Effect},'' {\em Phys. Rev.} {\bf D91} (2015) 045030,
\href{http://www.arXiv.org/abs/1407.1252}{{\tt 1407.1252}}.

\bibitem{Bergshoeff:2014uea}
E.~A. Bergshoeff, J.~Hartong, and J.~Rosseel, ``{Torsional Newton-Cartan
  geometry and the Schr\"odinger algebra},'' {\em Class. Quant. Grav.} {\bf 32}
  (2015), no.~13, 135017,
\href{http://www.arXiv.org/abs/1409.5555}{{\tt 1409.5555}}.

\bibitem{Bergshoeff:2017dqq}
E.~Bergshoeff, A.~Chatzistavrakidis, L.~Romano, and J.~Rosseel,
  ``{Newton-Cartan Gravity and Torsion},'' {\em JHEP} {\bf 10} (2017) 194,
\href{http://www.arXiv.org/abs/1708.05414}{{\tt 1708.05414}}.

\bibitem{Festuccia:2016awg}
G.~Festuccia, D.~Hansen, J.~Hartong, and N.~A. Obers, ``{Torsional
  Newton-Cartan Geometry from the Noether Procedure},'' {\em Phys. Rev.} {\bf
  D94} (2016), no.~10, 105023,
\href{http://www.arXiv.org/abs/1607.01926}{{\tt 1607.01926}}.

\bibitem{Bekaert:2014bwa}
X.~Bekaert and K.~Morand, ``{Connections and dynamical trajectories in
  generalised Newton-Cartan gravity I. An intrinsic view},'' {\em J. Math.
  Phys.} {\bf 57} (2016), no.~2, 022507,
\href{http://www.arXiv.org/abs/1412.8212}{{\tt 1412.8212}}.

\bibitem{banerjee:2016laq}
R.~Banerjee and P.~Mukherjee, ``{Torsional Newton–Cartan geometry from
  Galilean gauge theory},'' {\em Class. Quant. Grav.} {\bf 33} (2016), no.~22,
  225013,
\href{http://www.arXiv.org/abs/1604.06893}{{\tt 1604.06893}}.

\bibitem{Kaku:1978nz}
M.~Kaku, P.~K. Townsend, and P.~van Nieuwenhuizen, ``{Properties of Conformal
  Supergravity},'' {\em Phys. Rev.} {\bf D17} (1978)
3179.

\bibitem{Kaku:1978ea}
M.~Kaku and P.~K. Townsend, ``{Poincar\'e Supregarvity as Broken Superconformal
  Gravity},'' {\em Phys. Lett.} {\bf B76} (1978)
54.

\bibitem{Ferrara:1977ij}
S.~Ferrara, M.~Kaku, P.~K. Townsend, and P.~van Nieuwenhuizen, ``{Gauging the
  Graded Conformal Group with Unitary Internal Symmetries},'' {\em Nucl. Phys.}
  {\bf B129} (1977)
125.

\bibitem{Kaku:1977pa}
M.~Kaku, P.~K. Townsend, and P.~van Nieuwenhuizen, ``{Gauge Theory of the
  Conformal and Superconformal Group},'' {\em Phys. Lett.} {\bf B69} (1977)
304--308.

\bibitem{Niederer:1972zz}
U.~Niederer, ``{The maximal kinematical invariance group of the free
  Schrodinger equation.},'' {\em Helv. Phys. Acta} {\bf 45} (1972)
802--810.

\bibitem{Jackiw:1990mb}
R.~Jackiw and S.-Y. Pi, ``{Classical and quantal nonrelativistic Chern-Simons
  theory},'' {\em Phys. Rev.} {\bf D42} (1990) 3500.
[Erratum: Phys. Rev.D48,3929(1993)].

\bibitem{Henkel:1993sg}
M.~Henkel, ``{Schrodinger invariance in strongly anisotropic critical
  systems},'' {\em J. Statist. Phys.} {\bf 75} (1994) 1023--1061,
\href{http://www.arXiv.org/abs/hep-th/9310081}{{\tt hep-th/9310081}}.

\bibitem{Nishida:2007pj}
Y.~Nishida and D.~T. Son, ``{Nonrelativistic conformal field theories},'' {\em
  Phys. Rev.} {\bf D76} (2007) 086004,
\href{http://www.arXiv.org/abs/0706.3746}{{\tt 0706.3746}}.

\bibitem{Gibbons:2003rv}
G.~W. Gibbons and C.~E. Patricot, ``{Newton-Hooke space-times, Hpp waves and
  the cosmological constant},'' {\em Class. Quant. Grav.} {\bf 20} (2003) 5225,
\href{http://www.arXiv.org/abs/hep-th/0308200}{{\tt hep-th/0308200}}.

\bibitem{Grosvenor:2017dfs}
K.~T. Grosvenor, J.~Hartong, C.~Keeler, and N.~A. Obers, ``{Homogeneous
  Nonrelativistic Geometries as Coset Spaces},'' {\em Class. Quant. Grav.} {\bf
  35} (2018), no.~17, 175007,
\href{http://www.arXiv.org/abs/1712.03980}{{\tt 1712.03980}}.

\bibitem{ZOJER:2016aoj}
T.~Zojer, {\em {Non-relativistic supergravity in three space-time dimensions}}.
\newblock PhD thesis, Groningen U.,
2016-01-07.
\newblock

\bibitem{Andringa:2013mma}
R.~Andringa, E.~A. Bergshoeff, J.~Rosseel, and E.~Sezgin, ``{3D Newton-Cartan
  supergravity},'' {\em Class. Quant. Grav.} {\bf 30} (2013) 205005,
\href{http://www.arXiv.org/abs/1305.6737}{{\tt 1305.6737}}.

\bibitem{work-in-progress}
{Work in progress}.

\end{thebibliography}\endgroup

\end{document}